\documentclass[aps,prb,twocolumn,showpacs,preprintnumbers,amsmath,amssymb]{revtex4}
\usepackage{graphicx}
\usepackage{amsmath}	
\usepackage{bm}
\usepackage[usenames]{color}
\begin{document}
\draft
\title{
Optical manipulation of magnetism in spin-charge coupled correlated electron system
}
\author{Jun~Ohara$^{1,2, \dagger}$, Yu~Kanamori$^1$, and Sumio Ishihara$^{1,2}$} 
\address{$^1$Department of Physics, Tohoku University, Sendai 980-8578, Japan}
\address{$^2$Core Research for Evolutional Science and Technology (CREST), Sendai 980-8578, Japan}       
         
\date{\today}

\begin{abstract}
Photoirradiation effects in correlated electrons coupled with localized spins are studied 
based on the extended double-exchange model. 
In particular, we examine melting of an antiferromagnetic (AFM) charge order insulating state by varying the light intensity. 
When intense light is irradiated, the AFM insulating characteristics are strengthened, rather than change into the ferromagnetic metallic characteristic, which are expected from the conventional double exchange interaction when carriers are introduced by weak light irradiation or chemical doping. 
This provides a new principle for optically manipulating magnetism.
\end{abstract}

\pacs{78.47.J-, 75.78.Jp, 71.30.+h, 71.10.-w}
\maketitle

\section{Introduction}
\label{S:INTRO}
Optical manipulation of magnetism is one of the recent attractive themes in not only condensed matter physics but also modern nano-technologies. 
Since electron spins do not couple directly to the electric-field component of light, optical control of magnetism has been done in usual by utilizing the circular polarized light through the spin-orbit coupling. This strategy is applied successfully to the controls of magnetism in metallic magnets and some magnetic semiconductors. Another route for the optical control of magnetism is recently developed in correlated electron systems, in which magnetism is strongly coupled with electric transport. There is a possibility to change a magnetic state by modifying the charge state which can be directly controlled by light. One of the advantageous in this manner is a possibility of the ultrafast control of magnetism, since motion of electronic charge is much faster than pure spin motion which is usually governed by the exchange interaction. 

Such situation of strong correlation between magnetism and electric transport is realized in a system where local magnetic moments are embedded in a metallic solid.
Through the Hund's rule coupling between the local moments and the conduction electrons, 
the ferromagnetic (FM) double-exchange (DE) interaction operates between the spatially separated local moments.~\cite{zener, anderson}
Materials where the DE interaction gives rise to metallic ferromagnetism include magnetic semiconductors,~\cite{munekata,dietl}
colossal-magnetoresistive (CMR) manganites,~\cite{daggoto,tokura} and so on. 
The essence of the DE interaction is the kinetic-energy gain of conducting carriers; when local magnetic moments are parallel, electrons can hop between sites without reducing the Hund's rule coupling  energy.  

\begin{figure}[t]
\centering
\includegraphics[width=60mm]{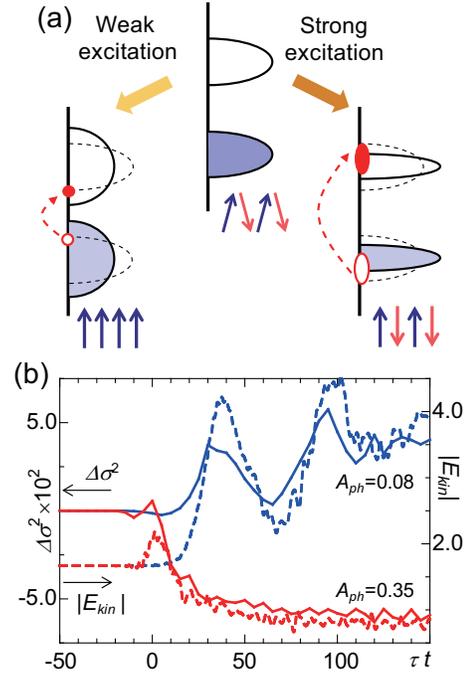}
\caption{(color online)
(a): Schematic diagram of changes in the electronic density-of-state (DOS) and magnetic structure 
on weak and strong photon excitations. 
Bold arrows represent localized spins. 
Broken arrows imply electronic excitations due to photoirradiation. 
(b): Changes in the band width $\Delta \sigma^2$ (bold lines) and the absolute value of the kinetic energy $E_{kin}$ (broken lines) as functions of time
for weak pumping $(A_{ph}=0.08)$ and strong pumping $(A_{ph}=0.35)$. }
\label{fig:kinetic}
\end{figure}
%
A FM metal originating from the DE interaction often competes with other electronic phases. 
In particular, competition between a FM metal and a charge-ordered (CO) antiferromagnetic (AFM) insulator is widely acknowledged to be the source of the CMR effects
in perovskite manganites.~\cite{tokura}
A CO AFM insulator is characterized by an energy gap due to the Coulomb interaction and
an alternating spin alignment by the AFM exchange interaction.
Hole doping by chemical substitution of a CO AFM insulator induces metallic conduction with a FM spin alignment. 
Many recent experimental and theoretical studies~\cite{fiebig,liu,ehrke,matsueda,koshibae} have revealed that these correlated changes are also realized by photo-doping that introduces carriers by optical laser pulses, as shown in the left of Fig.~\ref{fig:kinetic}(a).
This is a naive extension of the DE interaction to excited states, 
since photon pumping has been widely recognized to convert insulators to metals in a number of organic salts,~\cite{kawakami,chollet} transition-metal oxides~\cite{cavalleri,okamoto} and other correlated electron systems.~\cite{matveev,tsuji,moritz,eckstein}

In this paper, we demonstrate that this conventional mechanism for the DE interaction does not operate in highly photoexcited states. 
As shown in Fig.~\ref{fig:kinetic}(b), when intense light irradiation introduces carriers to an AFM insulator, the AFM insulating characteristics are strengthened, rather than change into the FM metallic characteristic, which are expected when it is irradiated by weak light. 
This provides a new principle for optically manipulating magnetism and conducting properties based on varying the light intensity.

\section{Model Hamiltonian and Methods}
\label{S:METHOD}
To examine changes in magnetism and conduction due to intense light irradiation, we adopt the extended DE model in which mobile electrons couple with local spins through FM Hund's rule coupling $J_H$.  
The explicit Hamiltonian is given by 
%
%
\begin{align}
{\cal H}_0=&-t\sum_{\langle ij \rangle , \sigma} ( c_{i \sigma}^\dagger c_{j \sigma} +  H.c. )
-J_H\sum_i {\bm s}_i \cdot {\bm S}_i
\nonumber \\
&
+J \sum_{\langle ij \rangle} {\bm S}_i \cdot {\bm S}_j
+V\sum_{\langle ij \rangle} n_i n_j,
\label{eq:hamiltonian}
\end{align}
where $c_{i \sigma}$ is the annihilation operator for an electron at site $i$ and spin $\sigma(=\uparrow, \downarrow)$, 
${\bm s}_i(\equiv \sum_{s, s'} c^\dagger_{i s} {\bf \sigma}_{ss'} c_{i s'} )$ and 
$n_i(\equiv \sum_\sigma n_{i \sigma}=\sum_\sigma c^\dagger_{i \sigma}c_{i \sigma} )$ 
are respectively the spin and charge operators for electrons, and ${\bm S}_i$ is the local spin operator. 
We restrict the present case to quarter-filling
so that the average number of mobile-electrons per site is 0.5. 
The first two terms correspond to the standard DE Hamiltonian which favors a FM metal. 
The competing phase, a CO AFM insulator, is stabilized by the Coulomb interaction between nearest neighbor (NN) sites, $V$, and the AFM exchange interaction between NN local spins, $J$. 
In the limit of strong Hund's rule coupling, $J_H \gg t$, an electron spin and a local spin at the same site are always parallel. 
This is performed by introducing the rotating frame at each site for the spin space of $c_{i\sigma}$.~\cite{korenman,tatara}
As a result, the conduction electrons are replaced by the spin-less fermions, $\widetilde c_i$, 
in which the hopping amplitudes depend on the local-spin directions.
Then, the original Hamiltonian is reduced to the following spin-less fermion model:  
\begin{align}
{\cal H}(\tau)=\sum_{\langle ij \rangle}
\left ( -{\widetilde t}_{ij} \widetilde c_i^\dagger \widetilde c_{j}+H.c.
+V \widetilde n_i  \widetilde n_j
+J {\bm S}_i \cdot {\bm S}_j \right ),
\label{eq:effectiveh}
\end{align}
where 
$\widetilde n_i=\widetilde c_i^\dagger \widetilde c_i$ is the number of spin-less fermions,
and 
the effective hopping amplitude for fermions depends on the angle of the local spins as~\cite{anderson}
\begin{align}
{\widetilde t}_{ij}=&t e^{ -i \int_{x_i}^{x_j} {\bm A}(\tau) \cdot d{\bm r} }
\{
e^{i(\phi_i-\phi_j)/2} \cos (\theta_i/2) \cos (\theta_j/2)
\nonumber \\
&+e^{-i(\phi_i-\phi_j)/2} \sin (\theta_i/2) \sin (\theta_j/2) \}.
\label{eq:efftrns}
\end{align}
Here the pump photon is introduced into Eq.~(\ref{eq:efftrns})
as the Peierls phase,
where ${\bm A}(\tau)$ is the vector potential for the pump photon at time $\tau$, 
and $x_i$ is a position of a site $i$. 
We assume that
${\bm A}(\tau)=A_{ph} {\bf \hat e} e^{-\gamma_0^2 \tau^2/2} \cos \omega_{ph} \tau$
with an amplitude $A_{ph}$, a damping factor $\gamma_0=0.1t$, a photon frequency $\omega_{ph}$, 
and a unit polarization vector ${\bf \hat e}$. 
The time origin is taken to be the center of the photon pulse. 
The pump photon intensity is expressed by a dimensionless vector-potential amplitude $A_{ph}$ in terms of the light velocity, the Planck constant and the lattice constant. 
The present values for an ''effective" absorbed photon density,~\cite{kanamori1}
$0 \le n_{ph} \le 0.15$ for $0 \le A_{ph} \le 0.35 $, are realistic in the usual optical pump-probe experiments.~\cite{matsubara}

The initial-state electronic structure at $\tau t\ll0$, time evolution, and excitation spectra are calculated by the numerical exact-diagonalization method in finite size clusters. 
The one- and two-dimensional $N$-site clusters
with the open-boundary condition are adopted. 
The mobile electron number is taken to be $N_{ele}=(N+1)/2$
which corresponds to the quarter filling in the case of the open boundary condition. 
The time evolution for the electronic wave function, $|\psi(\tau) \rangle$, is obtained by 
solving the time-dependent Schr$\ddot {\rm o}$dinger equation,
\begin{align}
i \frac{d | \psi (\tau) \rangle}{d \tau}  ={\cal H} (\tau)  | \psi (\tau) \rangle,
\label{eq:sheq}
\end{align}
where ${\cal H} (\tau)$ is the time-dependent Hamiltonian in Eq.~(\ref{eq:effectiveh}), 
based on the Lanczos method~\cite{park} (see Appendix~\ref{A:time}).
The pump photons are non-perturbatively taken into account.~\cite{ono, tanaka}
We confirmed that this numerical approach correctly reproduces
the exact analytic solution
describing the quantum tunneling phenomenon induced by the time-dependent
external field.~\cite{llorente, kayanuma}
The details are shown in Appendix~\ref{A:time}. 
As for the local spin sector, we introduce the classical equations of motion given by
\begin{align}
\frac{d}{d \tau}\frac{\partial  {\cal L}}{\partial \phi_i} -\frac{\partial {\cal L}}{\partial \dot \phi_i}=0
\label{eq:oleq1}
\end{align}
and
\begin{align}
\frac{d}{d \tau}\frac{\partial  {\cal L}}{\partial \theta_i}-\frac{\partial {\cal L}}{\partial \dot \theta_i}=0, 
\label{eq:oleq2}
\end{align}
where ${\cal L}$ is the Lagrangian defined by 
${\cal L}=S \sum_i \cos \theta_i \dot \phi_i -\langle {\cal H}(\tau) \rangle$,~\cite{obata}
and $\langle \cdots \rangle$ represents the average in terms of the wave function at time $\tau$.
For simplicity, magnitude of spin is assumed to be $S=1/2$. 
The equations of motion are solved by the fourth order Runge-Kutta method. 
In the numerical simulation,
we confirmed that the total energy after turnoff of the pumping is conserved
within 5$\times$10$^{-3}$ percent. 

\section{Photoinduced dynamics}
\label{S:DYNAMICS}
\begin{figure}[t]
\centering
\includegraphics[width=65mm]{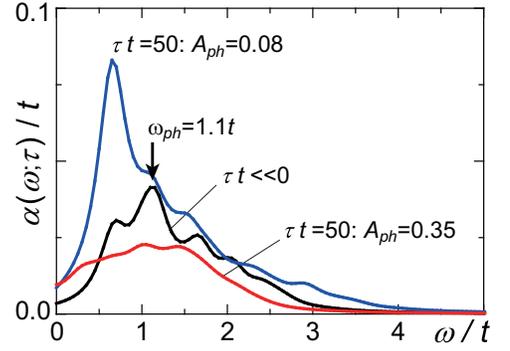}
\caption{
(color online): 
Transient optical absorption spectra after weak photon pumping ($A_{ph}=0.08$)
and strong photon pumping ($A_{ph}=0.35$)
at time $\tau t=50$. 
Absorption spectra for the initial CO-AFM insulator are also shown. 
Bold arrow indicates the pump photon energy. 
We note that transient spectra in the region of $|\omega|<0.02t$ are not meaningful because of the uncertainly principle between time and energy. }
\label{fig:oc}
\end{figure}

The ground-state phase diagram almost reproduces the previous results where the extended DE Hamiltonian is analyzed by the density-matrix renormalization-group (DMRG) method.~\cite{garcia,kanamori1}
We chose, as an initial state of photo-irradiation processes,  the spin-canted CO insulating phase which is located around a boundary between the CO-AFM insulating and FM metallic phases. 
The photon frequency $\omega_{ph}$ is tuned to the insulating gap energy (see an arrow in Fig.\ref{fig:oc}).

\begin{figure}[t]
\centering
\includegraphics[width=86mm]{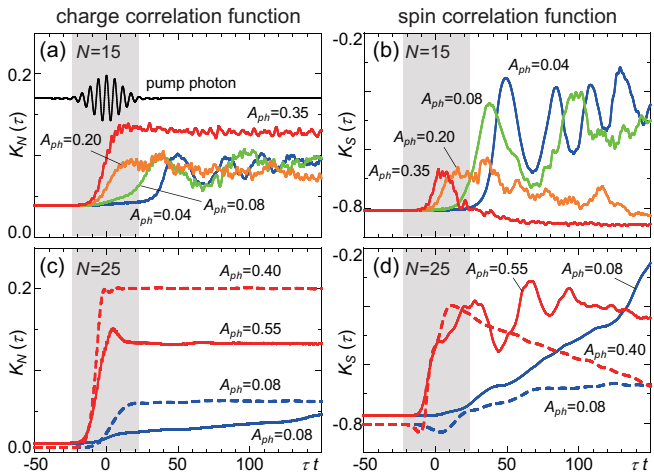}
\caption{
(color online): 
Time profiles of the charge and spin correlation functions
for several pump-photon amplitudes. 
Results for the $N=15$ chain cluster where $V/t$=1.4 and $J/t$=1.24 are shown in (a) and (b).
The pump-photon profile with $\omega_{ph}=1.1t$ is also shown,
and photon amplitudes are $A_{ph}=0.04,\ 0.08,\ 0.2,\ 0.35$.
Results for the $N=25$ chain (solid lines: $V/t=3.1, J/t=0.625$) and
$N=5 \times 5=25$ square (broken lines: $V/t=2.1, J/t=0.31$) clusters are shown in (c) and (d).
Pump photon energies are chosen to be 2.7$t$ and 6.2$t$ for the chain and square clusters, respectively.}
\label{fig:correlation}
\end{figure}
Real-time spin- and charge-dynamics after photon-pumping were respectively monitored by the correlation function between the NN localized spins, 
$K_S=N_B^{-1}S^{-2}\sum_{\langle ij \rangle} {\bm S}_i \cdot {\bm S}_j$, 
and that between NN mobile charges, 
$K_N=N^{-1}_B\sum_{\langle ij \rangle}\langle {\widetilde n}_i {\widetilde n}_j \rangle$, where $N_B$ is the total number of NN bonds. 
Time evolutions of $K_S(\tau)$ and $K_{N}(\tau)$ for several photon amplitudes calculated
in a $N=15$ chain are shown in Figs.~\ref{fig:correlation}(a) and ~\ref{fig:correlation}(b). 
The elapsed time $\tau$ is denoted in units of $t^{-1}$; it is a few femto-seconds for conventional transition-metal oxides. 
In an ideal CO-AFM insulator, we have $K_N=0$ and $K_S=-1$.
The increase in $K_N$ with $A_{ph}$ implies that the initial coherent CO collapses on irradiation. 
In contrast, the time profiles for $K_S$ are rather complex. 
For weak pumping $(A_{ph} \lesssim 0.08)$, the AFM correlations gradually weaken, and exhibit oscillating behavior
[see the green line in Fig.~\ref{fig:correlation}(b)  for example]. 
This oscillation might be attributed to the Rabi oscillation between the photoexcited states.
These profiles in $K_S$ are well understood from the conventional DE mechanism.~\cite{matsueda,kanamori2}
%
In contrast, for strong pumping $(0.2 \lesssim A_{ph})$,
the AFM correlations suddenly weaken after photon pumping, and gradually recover to the initial AFM state
[see the red line in Fig.~\ref{fig:correlation}(b) for example].
After $\tau t \sim 35$, the AFM correlation is slightly stronger than its initial value.  
In other words, for $K_S$ when $\tau  t \gtrsim  35$, fewer photons collapse the AFM correlation more effectively. 
These characteristic features in $K_S(\tau)$ and $K_{N}(\tau)$ presented are robust by changing the clusters,
as shown in Figs.~\ref{fig:correlation}(c) and \ref{fig:correlation}(d),
where the blue and red lines indicate the weak and strong pumping cases, respectively.
Hereafter, we focus on the dynamics in the $N=15$ chain to avoid redundancy and complexity.

To investigate change in the carrier motion predicted by the conventional DE mechanism,
Fig.~\ref{fig:kinetic}(b) shows time profiles for the kinetic energy of the mobile carriers,  
$E_{kin} \equiv \langle - \sum_{\langle ij \rangle} {\widetilde t_{ij}} \widetilde c_{i}^\dagger \widetilde c_{j} +H.c. \rangle$ (see broken lines). 
For weak pumping ($A_{ph}=0.08$), $|E_{kin}|$ increases accompanied with oscillations, as expected from the DE mechanism.
Completely different behavior is observed for strong pumping ($A_{ph}=0.35$): $|E_{kin}|$ decreases monotonically after pumping. %
These results indicate that this highly excited state is a charge-disordered AFM insulator.  
Since the total energy is conserved after turning off the pump photon,
the kinetic-energy gain after weak pumping is compensated by a loss in the AFM exchange-energy of the local spins
[see the blue broken line in Fig.~\ref{fig:kinetic}(b) and the green solid line in Fig.~\ref{fig:correlation}(b)]. 
In contrast, after strong pumping, reduction(s) in the kinetic energy is compensated by gain in the AFM exchange-energy due to reinforcement of the AFM correlation
[see the red broken line in Fig.~\ref{fig:kinetic}(b) and the red solid line in Fig.~\ref{fig:correlation}(b)].

\begin{figure}[t]
\centering
\includegraphics[width=86mm]{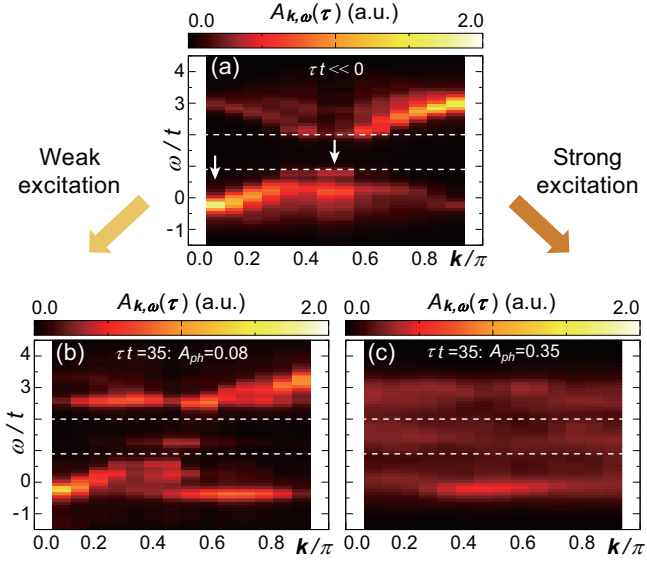}
\caption{(color online)
Contour plots of one-particle excitation spectra in the momentum-energy space. 
(a) Results for the initial CO-AFM insulator at $\tau t \ll 0$. 
White arrows denote the top and bottom of the valence band.
(b) The spectra after weak photon pumping $(A_{ph}=0.08)$, 
and (c) the spectra after strong photon pumping $(A_{ph}=0.35)$ at $\tau t=$35. 
White broken lines represent the top of the valence band and the bottom of the conduction band in the initial insulating state. 
We note that transient spectra in the region of $|\omega-1.4t|<0.03t$ are not meaningful because of the uncertainly principle between time and energy. }
\label{fig:onep}
\end{figure}
Time- and momentum-resolved electronic structures provide insights into what triggers these unusual highly photoexcited states. 
Transient correlated electronic bands are monitored by
time-dependent one-particle excitation spectra derived from the retarded Green's function.
This is divided by the electronic part defined by 
\begin{align}
A^{e}_{{\bm k}, \omega} (\tau)=-\frac{1}{\pi}
{\rm Im} \sum_{n m l}
\frac{
\psi_n^\ast(\tau)
({\widetilde c}^\dagger_{\bm k})_{nm}
({\widetilde c_{\bm k}})_{ml}
\psi_l(\tau)}{
(-\omega-E_m+E_n+i \eta)}
\label{eq:aqwele}
\end{align}
and the hole part by 
\begin{align}
A^{h}_{{\bm k}, \omega}(\tau)=-\frac{1}{\pi}
{\rm Im}\sum_{n m l}
\frac{
\psi_n^\ast(\tau) 
({\widetilde c}_{\bm k})_{nm}
({\widetilde c}^\dagger _{\bm k})_{ml}
\psi_l(\tau)}{
(\omega-E_m+E_{l}+i \eta)},
\label{eq:aqwhole}
\end{align}
where we introduce the eigen energy $E_m$, the eigen state $|m \rangle $, 
$({\widetilde c}_{\bm k})_{nm}=\langle n |{\widetilde c}_{\bm k}| m \rangle$, 
$\psi_n(\tau)=\langle n|\psi(\tau) \rangle$, and an infinitesimal constant $\eta=0.1t$.
These formulae are derived based on the linear response theory 
for a probe photon in the photoexcited transient state.~\cite{kanamori1}
We show the derivation of Eqs.~(\ref{eq:aqwele}) and (\ref{eq:aqwhole}) 
and their validity in Appendix B.
Results for weak and strong pumping are presented in Fig.~\ref{fig:onep}. 
In the initial CO-AFM insulator (see Fig.~\ref{fig:onep}(a)), a direct gap in the electron energy bands opens at the center of the Brillouin zone, e.g. ${\bm k}= \pi/2$. 
After photon-pumping, a large change in the electronic spectra occurs due to many-body effects. 
%
For weak pumping (see Fig.~\ref{fig:onep}(b)), changes in the electronic structure are caused by the spectral-weight transfer around ${\bm k}=\pi/2$ from the valence band to the in-gap region. 
Specifically, electrons at the top of the valence band are excited, and a new in-gap state is generated around the same momentum. 
On the other hand, for strong pumping (see Fig.~\ref{fig:onep}(c)), the electronic-structure changes across the Brillouin zone.
In particular, high spectral weights around the bottom of the valence band (${\bm k}\simeq0$) in the initial state are smeared out, 
and are transferred to the initial gap region and the whole valence-band region. %

To analyze the transient valence-band width in more detail, 
we introduce the second moment of the valence band defined by
\begin{align}
\sigma^2(\tau)=\frac{1}{N} \int_{- \infty}^{E_{top}} \sum_{\bm k} 
A^{e}_{{\bm k}, \omega} (\tau) (\omega-\omega_c)^2  d \omega,
\label{eq:disp}
\end{align}
where $\omega_c$ is the center of the band given by 
\begin{align}
\omega_c=
\frac{
\int_{- \infty}^{E_{top}} \sum_{\bm k} 
A^{e}_{{\bm k}, \omega} (\tau) \omega  d \omega}
{\int_{- \infty}^{E_{top}} \sum_{\bm k} 
A^{e}_{{\bm k}, \omega} (\tau)  d \omega}
\label{eq:omegac}
\end{align}
and $E_{top}$ is the energy at the top of the valence band for $\tau t \ll 0$. 
We present the difference from the initial band width as 
$\Delta \sigma^2=\sigma^2(\tau)-\sigma^2(\tau \ll 0)$ in Fig.~\ref{fig:kinetic}(b).
For weak pumping ($A_{ph}=0.08$),
$\Delta \sigma^2$ increases and shows positive values, as expected from the conventional DE scenario.
On the other hand, results are completely different for strong pumping ($A_{ph}=0.35$). 
The band width becomes wide (narrow) for weak (strong) pumping. 
These are the similar behaviors with the changes in the kinetic energy $E_{kin}$ calculated directly. 

We now present a physical picture for the highly photoexcited state (see Fig.~\ref{fig:kinetic}(a)). 
For weak pumping, as shown in Fig.~\ref{fig:onep}(b),
holes are generated near the top of the valence band. This situation is similar to the state
induced by chemical doping of holes. 
According to the conventional DE mechanism, 
a broadening of the valence band accompanied with the FM alignment of local spins occurs to gain the hole kinetic-energy.
However, for strong pumping, as shown in Fig.~\ref{fig:onep}(c), many photodoped holes are introduced near the bottom of the band. 
We consider that the multi-photon processes in the strong light intensity cause the high-energy excitations.
The induction of such high-energy states is observed as the large-scale changes in
the one-particle excitation spectra just after the pumping.
The band narrowing increases the hole kinetic-energy gain. 
This highly photoexcited state is the origin of the enhanced AF insulator characteristics induced by strong photon pumping. 

We propose that time-resolved pump-probe experiments can confirm the present unusual photoexcited state. 
The pump-probe optical absorption spectra is calculated by the linear response theory for the probe photon defined by 
\begin{align}
\alpha(\omega ; \tau)=-\frac{1}{\pi N}
{\rm Im} 
\sum_{nml}
\frac{
\psi_n^\ast(\tau)
({\cal J})_{nm}({\cal J})_{ml} \psi_l(\tau)}
{(\omega-E_m+E_{l}+i \eta)},
\label{eq:oc}
\end{align}
where we introduce the current operator given by 
$
{\cal J}=-i \sum_{\langle ij \rangle}  {\widetilde t}_{ij}
{\widetilde c}_i^\dagger {\widetilde c}_j + H.c.$ (Ref.~\onlinecite{kanamori1})
and 
$({\cal J})_{nm}=\langle n | {\cal J} |m \rangle$. 
As shown in Fig.~\ref{fig:oc}, a gap-like feature is observed in $\alpha(\omega; \tau)$
for an initial CO insulator at $\tau t \ll 0$. 
A low-energy peak grows in the initial optical gap for $A_{ph}=0.08$, as expected for the metallic state. 
On the other hand, the reduction in the whole spectral intensity in $A_{ph}=0.35$ implies enhancement of the localized character of mobile carriers. 
The low spectral intensity in the initial gapped region for strong pumping indicates collapse of the coherent CO state. 

\section{Discussion and Summary}
\label{S:DISCUSSION}
We reveal a new aspect of the DE system. The reinforcement of the AFM correlation and the band narrowing observed in the present study were not expected for chemical doping and thermal carrier creation.
These phenomena are only realized in highly photoexcited states and 
will be directly detected by time-resolved resonant x-ray diffraction and angular resolved PES.
After several-hundred femto seconds, which is beyond the limits of the present simulation, spin relaxation originating from the spin-orbit interaction begins to occur and it tends to stabilize the FM metal. 
Energy relaxation will eventually restore the system to the initial CO-AFM insulating state.  
In other words, relaxation effects for both angular momentum and energy relax the system to trivial fixed points. 
The present unique photo-induced phenomena occur in early stages before relaxation starts to operate.
However, they have been overlooked in the experimental observation until now. 
This is due to a prejudice based on the conventional DE scenario,
i.e. the light-irradiation is believed to always bring(s) about a FM metallic state.
We expect the precise optical pump-probe measurements
focusing on the immediate dynamics after strong photopumping
(within several tens of femtoseconds),
especially in the perovskite-type manganese oxides.

We suggest that similar photo-induced phenomena occur in several systems where the band width is directly controlled by spin, orbital and lattice degrees of freedom. 
It is proposed that a photo-induced hidden state will be searched at an early stage after pumping by changing the light intensity. 
The present study does not only demonstrate a route to the optical manipulation of magnetism but also provide a way to acquire a new state of matter by means of light. 

\acknowledgements
We thank H.~Matsueda, S.~Koshihara, S.~Iwai, and Y.~Okimoto for helpful discussions.
This work was supported in part by KAKENHI, DYCE (Optical Science of Dynamically
Correlated Electrons, 23104703). Some of the numerical calculations were performed using the supercomputing facilities at ISSP, the University of Tokyo, and Kyoto University. 

\noindent
$^{\dagger}$Present address: 
Department of Physics, Hokkaido University, Sapporo 060-0810, Japan.

\appendix
\section{Time-evolutional Calculation}
\label{A:time}

In this Appendix, we present a detailed formulation to solve the time-dependent 
Schr$\ddot{\rm o}$dinger equation for the time-dependent Hamiltonian given in Eq.~(\ref{eq:sheq}) based on the Lanczos method. 
We also show an application of this method to a two-level system in order to show the validity. 

In Eq.~(\ref{eq:sheq}), the time-dependent wave function is expanded as
\begin{equation}
| \psi(\tau) \rangle=\sum_i a_i(\tau) | \phi_i \rangle,
\label{eq:psi}
\end{equation}
where $\{ | \phi_i \rangle \} $ is taken to be a time-independent ortho-normal basis set.
Although it is possible to introduce the time-dependent phase factor to the basis set
as $| \phi_i \rangle \rightarrow | \phi_i(\tau) \rangle$,
this can be included in the coefficient $a_i(\tau)$ by changing its definition.
Through the matrix equation of motion for the basis set $\hat a(\tau) \equiv \{a_i(\tau)\}$ given by 
\begin{equation}
i \frac{d \hat a(\tau)}{d \tau}=\hat {\cal H}(\tau) \hat a(\tau), 
\end{equation}
we have 
\begin{equation}
 \hat a(\tau)={\it T} \exp \left [-i \int_0^\tau \hat {\cal H}(\tau')d\tau' \right ] \hat a(0) , 
\label{eq:at}
\end{equation}
where $\it T$ is the time-ordering operator, and $\hat {\cal H} (\tau)$ is the matrix formula of ${\cal H}(\tau)$. 
In the numerical calculation, the integral in Eq.(\ref{eq:at}) is performed step by step during 
the infinitesimal time, $\delta \tau$, as 
\begin{equation}
\hat a(\tau+\delta \tau)= \sum_{n=1}^{M}\exp (-i E_{n} \delta \tau )
\hat{\alpha}_{n} [\hat{\alpha}_{n}^{\dag} \cdot \hat a(\tau)],
\label{eq:adltt}
\end{equation}
where $E_{n}$ and $\hat{\alpha}_{n}$ are respectively
the pseudo eigenvalue and the pseudo eigenvector
obtained by the $M$-step Lanczos procedure for $\hat{{\cal H}}(\tau)$
starting from the trial function $\hat a(\tau)$.
The above formula for the time-dependent Schr\"{o}dinger 
equation for the time-dependent Hamiltonian is exact within
the numerical errors arising from the approximations that
(i) the Lanczos iterations are taken to be $M$ which are smaller than the Hilbert space dimensions, 
and (ii) the integral in Eq.~(\ref{eq:at}) is replaced by the successive time evolution
during a small time interval $\delta \tau$. 
We set $M=20$ and $\delta \tau=10^{-3}t^{-1}$ in the present calculations 
and checked sufficient accuracy.

\begin{figure}[h]
\centering
\includegraphics[width=60mm]{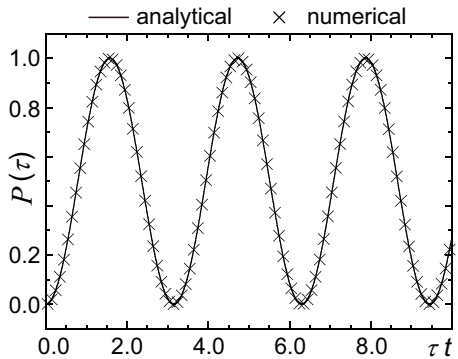}
\caption{(color online):
Time profiles of $P(\tau)$ obtained by the analytical (solid line)
and numerical (symbols) solutions of the two-level system,
where the parameters are set to be $E_{0}/t=10$ and $\omega/t=100$.
We chose $M=2$ and $\delta \tau=10^{-1}t^{-1}$ in the numerical calculation.}
\label{F:RATIO}
\end{figure}

In order to show validity of this method, we present here a simple application of this method to the two-level system:~\cite{llorente, kayanuma}
\begin{equation}
{\cal H}_{\rm two}(\tau)=\frac{E_0}{2} \cos(\omega \tau)  \Bigl(|1 \rangle \langle 1|- |2 \rangle \langle 2| \Bigr ) 
+t \Bigl (|1 \rangle \langle 2|+ |2 \rangle \langle 1| \Bigr ) , 
\label{eq:two}
\end{equation}
where $|1\rangle$ and $|2\rangle$ represent the left and right localized states, respectively, 
$(E_0/2) \cos(\omega \tau)$ is the time-dependent external field, and $t$ is the tunneling amplitude. 
We calculate $P(\tau)=| \langle 2 | \psi(\tau) \rangle |^2$ which
represents a probability that the system is in the state $|2 \rangle$ at time $\tau$. 
The initial state at $\tau=0$ is chosen as $|1 \rangle$.
The numerical solutions obtained by Eq.~(\ref{eq:adltt}) and the exact analytical solutions 
for $\omega/t \gg 1$ in Refs.~\onlinecite{llorente} and ~\onlinecite{kayanuma}
are presented in Fig.~\ref{F:RATIO}.
The numerical results almost completely reproduce the analytical results.

\section{Time-resolved excitation spectra}
\label{A:linear}

In this Appendix, 
we present a detailed formulation for the time-resolved excitation spectra,
given in Eqs. (\ref{eq:aqwele}), (\ref{eq:aqwhole}) and (\ref{eq:oc}). 

We assume that the external field in probe is week and 
apply a linear-response theory (LRT) to the states after photoexcitation.~\cite{kanamori1}
We first present a formulation of
the time-resolved optical-absorption spectra.
The probe photon is introduced by the uniform vector potential given as
\begin{align}
A_{probe}(\tau)=A_{probe}e^{-\gamma|\tau-\tau_{D}|-i\omega\tau}
\end{align}
with a probe-photon energy $\omega$, the center of an envelope $\tau_{D}$,
and a damping factor $\gamma$.
We assume that the probe pulse has no overlap with the pump pulse
and the pump photon is fully damped at time $\tau$.
These are given by the conditions
$\tau_{D}-\gamma^{-1}\gg \sqrt{2}\gamma_{0}^{-1}$ and $\tau\gg\sqrt{2}\gamma_{0}^{-1}$,
where $\gamma_{0}$ is the damping factor of the pump photon given in the Sec.~\ref{S:METHOD}. 
The interaction between the probe photon and the spinless fermions, 
introduced in Eq.~(\ref{eq:effectiveh}),
at time $\tau$ is defined by
\begin{align}
{\cal H}'(\tau)=-jA_{probe}(\tau),
\end{align}
where $j$ is the current operator.
The time evolution of the electronic wave function by the probe photon is given by
\begin{align}
|\tilde{\psi}(\tau) \rangle=
U_{0}(\tau,\tau_{0})U_{1}(\tau,\tau_{0})|\psi(\tau_{0}) \rangle,
\label{E:wave}
\end{align}
where
$|\psi(\tau_{0}) \rangle$ 
is the solution of the time-dependent Schr$\ddot{\rm o}$dinger equation in Eq.~(\ref{eq:sheq}) 
at time $\tau_{0}$ after turning off the pumping
($\tau_{D}-\gamma^{-1}\gg \tau_{0} \gg \sqrt{2}\gamma_{0}^{-1}$).
The time-evolution operators are given by
\begin{align}
U_{0}(\tau,\tau_{0})&=\exp{[-i{\cal H}(\tau-\tau_{0})]},
\label{E:u0} \\
U_{1}(\tau,\tau_{0})&=T\exp{[i\int_{\tau_{0}}^{\tau}d\tau'j(\tau',\tau_{0})A_{probe}(\tau')]},
\label{E:u1}
\end{align}
with
\begin{align}
j(\tau',\tau_{0})=U_{0}^{\dag}(\tau',\tau_{0})jU_{0}(\tau',\tau_{0}),
\label{E:j}
\end{align}
where ${\cal H}$ is given in Eq.~(\ref{eq:effectiveh}), but 
is no longer time-dependent operator, 
because the pump pulse is already turned off here.
Equation~(\ref{E:u1}) is expanded as
\begin{align}
U_{1}(\tau,\tau_{0})=1+i\int_{\tau_{0}}^{\tau}d\tau'j(\tau',\tau_{0})A_{probe}(\tau')+{\cal O}(A_{probe}^{2}).
\label{E:u1'}
\end{align}
Hereafter, ${\cal O}(A_{probe}^{2})$ term is ignored.
By using the wave function given in  Eq.~(\ref{E:wave}), 
the expectation value of the current operator is calculated as a 
product form of the response function and $A_{probe}(\tau)$. 
The former is obtained by the current-current correlation function given by
\begin{align}
-i\int_{\tau_{0}}^{\tau}d\tau'
\langle \psi(\tau_{0})|[j(\tau',\tau_{0}),j(\tau,\tau_{0})]|\psi(\tau_{0})\rangle
e^{-\gamma|\tau'-\tau_{D}|-i\omega\tau'},
\label{E:jj}
\end{align}
in common with the conventional LRT.~\cite{kanamori1}
The coefficient of $e^{-\gamma|\tau-\tau_{D}|-i\omega\tau}$ in Eq.~(\ref{E:jj})
corresponds to the optical absorption spectrum $\alpha(\tau, \omega)$ at time $\tau$,
and we obtain the form of Eq.~(\ref{eq:oc}).
The above formulae for the transient spectrum are valid in the region of $|\omega| > \tau^{-1}$
because of the uncertainly principle between time and energy.

   As with the optical absorption spectrum,
the one-particle excitation spectra are obtained
as a correlation function of the creation and annihilation operators by,
\begin{align}
-i\int_{\tau_{0}}^{\tau}d\tau'
\langle \psi(\tau_{0})|[c_{\bm k}^{\dag}(\tau',\tau_{0}),c_{\bm k}(\tau,\tau_{0})]|\psi(\tau_{0})\rangle
e^{-\gamma|\tau'-\tau_{D}|-i\omega\tau'},
\label{E:cc}
\end{align}
where $c_{\bm k}^{(\dag)}(\tau_{1},\tau_{2})=
U_{0}^{\dag}(\tau_{1},\tau_{2})c_{\bm k}^{(\dag)}U_{0}(\tau_{1},\tau_{2})$.
The coefficient of $e^{-\gamma|\tau-\tau_{D}|-i\omega\tau}$ in Eq.~(\ref{E:cc})
consists of the electron part $A^{e}_{{\bm k}, \omega} (\tau)$
and the hole part $A^{h}_{{\bm k}, \omega} (\tau)$.
These forms are given by Eqs.~(\ref{eq:aqwele}) and (\ref{eq:aqwhole}).
The transient one-particle excitation spectra for the present gapped system
are valid in the region of $|\omega-\omega_{g}|>\tau^{-1}$
because of the uncertainly principle between time and energy,
where $\omega_{g}$ is the center of the energy gap at $\tau t \ll 0$. 

\begin{figure}[h]
\begin{center}
\includegraphics[width=60mm]{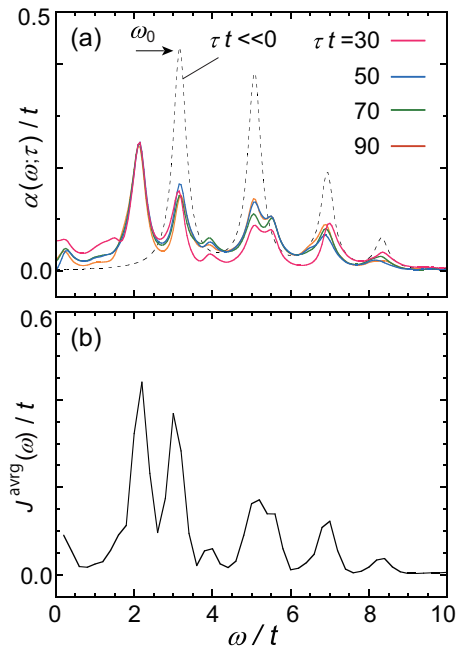}
\end{center}
\caption{(color online):
The optical responses in the one-dimensional $V$-$t$ model calculated by the two methods. 
(a) The optical absorption spectra calculated by LRT for the probe photon. 
The black dotted line indicates the absorption spectrum before pumping ($\tau t\ll0$).
(b) The current induced by the probe photon calculated by Eq.~(\ref{eq:avcrnt})
as functions of the probe photon frequency. 
The half-filled $N=9$ site cluster with an open-boundary condition is adopted. 
The parameters are set to be $V/t=5.0$, $A_{pump}=0.5$, and $\gamma_0=0.7t$.
The pump photon energy $\omega_{0}$ is taken to be $3.0t$ corresponding to the charge-gap energy in the initial state. 
Parameter values in (b) are chosen to be $A_{probe}=0.05$, $\gamma=0.2t$, and $\tau_{D}t=30$. 
We set $M=20$ and $\delta \tau=10^{-2}t^{-1}$ in both the time-evolutional calculations.} 
\label{F:SPECTRA}
\end{figure}

In order to demonstrate validity of the above formulae, 
we adopt them to the one-dimensional $V$-$t$ model defined by
\begin{align}
{\cal H}_{Vt}=-t
\sum_{i=1}^{N-1} (c_{i}^\dagger c_{i+1} + {\rm H.c.} )+ V\sum_{i=1}^{N-1} n_{i} n_{i+1},
\label{eq:vt}
\end{align}
where $c_i$ is the annihilation operator for the spinless fermion, and $n_i$ is the number operator. The pump photon is introduced as the Peierls phase. 
The optical responses in the photoexcited state are calculated 
by the two methods; 
after turning off the pump photon, 
we calculate (i) the optical-absorption spectra by using the formulae given above, 
and (ii) the electric current induced by the probe photon by using the following 
non-perturbative method.
In the latter, we introduce that both the pump and probe photons are the damped oscillator forms as 
\begin{align}
A(\tau)&=A_{pump}e^{-\gamma_0^2 \tau^2/2} \cos(\omega_0 \tau)
\nonumber \\ &
+ A_{probe} e^{-\gamma^2(\tau-\tau_{D})^{2}/2} \cos \{\omega(\tau-\tau_{D})\} , 
\end{align}
where $\omega_0$ ($\omega$) is the frequency 
and $\gamma_0$ ($\gamma$) is the damping constant for the pump (probe) photon.  
After turning off the pump photon,
the current at time $\tau$ induced by the probe photon is defined by
\begin{align}
J(\omega, \tau) = {\cal J}(\tau, A_{probe}=0.05,\omega)
                 -{\cal J}(\tau, A_{probe}=0,\omega), 
\label{eq:nlrt}
\end{align}
where the first and second terms represent the current with and without the probe photon. 
We define 
$
{\cal J}(\tau, A_{probe},\omega)=\langle \psi (\tau) |j| \psi(\tau) \rangle,
$
where $| \psi(\tau) \rangle$ is the wave function at time $\tau$ obtained from Eq.~(\ref{eq:sheq})
and $j$ is the current operator. 
We calculate a time-averaged $J(\omega, \tau)$ during time interval after the "probe" photon irradiation. 
given as
\begin{equation}
{\cal J}^{\rm avrg}(\omega)=\frac{1}{50}\int_{40}^{90}\sqrt{\bigl( {\cal J}(\omega,\tau) \bigr)^{2}}d(\tau t),
\label{eq:avcrnt}
\end{equation}
where the center of the probe-photon envelope is taken to be $\tau_{D} t=30$. 
This is interpreted as a response to the "probe" photon,
since the current induced by "probe" is not dissipated in the present system 
after turning off the probe.

   In Fig.~\ref{F:SPECTRA}, the numerical results of the optical conductivity spectra calculated by
LRT and the current induced by the probe photon are compared.  
In the results by LRT [Fig.~\ref{F:SPECTRA}(a)], the large peaks are seen in the transient optical spectra 
(colored curves) around $\omega/t =$ 2, 3, 5 and 7. 
The spectra calculated by the present non-perturbative method [Fig.~\ref{F:SPECTRA}(b)]
are well reproduced by the $\omega$-dependence of the optical conductivity calculated by LRT,
even for the small peaks around $\omega=$4 and 8.3,
and shoulder structures around $\omega=1.5$ and 5.5.

\end{document}